\documentclass[amsmath,amssymb]{arXiv}

\newcommand {\Fig}[1] {Fig.~\ref{#1}}
\newcommand{\sitweight}{$^{28}$Si}
\newcommand{\ptone}{$^{31}$P}
\newcommand{\dox}{D$\rm ^0$X}

\usepackage{color}
\usepackage[]{graphicx}

\title{Hybrid optical-electrical detection of donor electron spins with bound excitons in silicon}

\author{C. C. Lo$^{1,2}$, M. Urdampilleta$^1$, P. Ross$^1$, M. F. Gonzalez-Zalba$^3$, J. Mansir$^{1}$, S. A. Lyon$^4$, M. L. W. Thewalt$^5$ \& J. J. L. Morton$^{1,2}$}

\begin{document}

\maketitle

\begin{affiliations}
	\item London Centre for Nanotechnology, University College London, London WC1H 0AH, UK.
	\item Department of Electronic and Electrical Engineering, University College London, London WC1E 7JE, UK.
	\item Hitachi Cambridge Laboratory, J. J. Thomson Avenue, Cambridge CB3 0HE, U.K.
	\item Department of Electrical Engineering, Princeton University, Princeton, New Jersey 08544, USA.
	\item Simon Fraser University, Burnaby, British Columbia V5A 1S6, Canada.
\end{affiliations}

\begin{abstract}
Electrical detection of spins is an essential tool in understanding the dynamics of spins in semiconductor devices, providing valuable insights for applications ranging from optoelectronics\cite{malissa14, algasinger13}
and spintronics\cite{appelbaum07} to quantum information processing\cite{mccamey10, paik10, stegner06, lo11, lo13}. 
For electron spins bound to shallow donors in silicon, bulk electrically-detected magnetic resonance has relied on coupling to spin readout partners such as paramagnetic defects\cite{stegner06, paik10} or conduction electrons\cite{mccamey10, lo11, lo13} which fundamentally limits spin coherence times.
Here we demonstrate electrical detection of phosphorus donor electron spin resonance by transport through a silicon device, using optically-driven donor-bound exciton transitions\cite{steger12, saeedi13}. 
We use this method to measure electron spin Rabi oscillations, and, by avoiding use of an ancillary spin for readout, we are able to obtain long intrinsic electron spin coherence times, limited only by the donor concentration\cite{tyryshkin12}. 
We go on to experimentally address critical issues for adopting this scheme for single spin measurement in silicon nanodevices, including the effects of strain, electric fields, and non-resonant excitation. This lays the foundations for realising a versatile readout method for single spin readout with relaxed magnetic field and temperature requirements compared with spin-dependent tunneling\cite{morello10, pla12}.
\end{abstract}

\newpage

Shallow donor electron and nuclear spins in silicon have extraordinarily long coherence times\cite{tyryshkin12, steger12, saeedi13}, making them attractive candidates for quantum information processing\cite{kane98}, quantum memory\cite{morton08}, as well as for quantum sensing applications\cite{taylor08}. In addition, neutral shallow donors can form bound exciton states (D$\rm ^0$X) with relatively long lifetimes ($\sim$200--300~ns~[\citen{schmid77}]) and correspondingly narrow intrinsic line widths, enabling optical transitions with both electron and nuclear spin selectivity\cite{yang09}. D$\rm ^0$X can be formed by a direct photon excitation (a no phonon transition) with energy $E[{\rm D^0X}]\approx\:$1.15$\:$eV, below the silicon indirect bandgap of $E_g\:=\:$1.17$\:$eV
. The photon excites an electron from the valence band at the neutral donor site, resulting in two indistinguishable electrons and one hole localised in the \dox\ ground state. The \dox\ state then relaxes via an Auger recombination process where the excess electron-hole pair recombines and its energy is transferred to the remaining electron, ejecting it from the donor site and leaving behind the positively charged donor ion.

The D$^0$X spin-selective optical transitions are attractive for realising hybrid optical-electrical ensemble spin detection in silicon since no decoherence-inducing paramagnetic centre close to the dopant is required for spin-charge conversion. 
In addition, they could enable high-fidelity single dopant electron spin readout without the requirement of keeping the thermal energy much less than the Zeeman splitting  --- this is in contrast with spin-dependent tunneling schemes\cite{morello10}, where, for example, $\approx\:$80$\:\%$ electron spin readout fidelity was achieved at $T\:\approx\:$300$\:$mK and $B\:\approx\:$1$\:$T~[\citen{pla12}]. The maximum temperature for spin readout using D$\rm ^0$X is limited by its dissociation energy of approximately 5$\:$meV~[\citen{haynes60}], such that this could be readily implemented at liquid helium temperatures. 
Spin readout fidelity is instead determined by the optical transition linewidths (approximately 20$\:$neV when lifetime limited) compared to the D$\rm ^0$X spin splitting, which is always at least the hyperfine interaction strength ($A$~=~0.486~$\mu$eV for phosphorus donors (\ptone)). Hence, zero magnetic field measurements are possible in principle. 
These strongly relaxed experimental conditions open the possibility of practical implementation of quantum sensing applications with donor spins\cite{taylor08}, and enable access to the long donor spin relaxation times observed at low magnetic fields ($B\ll1$~T)~[\citen{tyryshkin12}]. 
 
 \begin{figure}
	\centering
	\includegraphics[width=16cm]{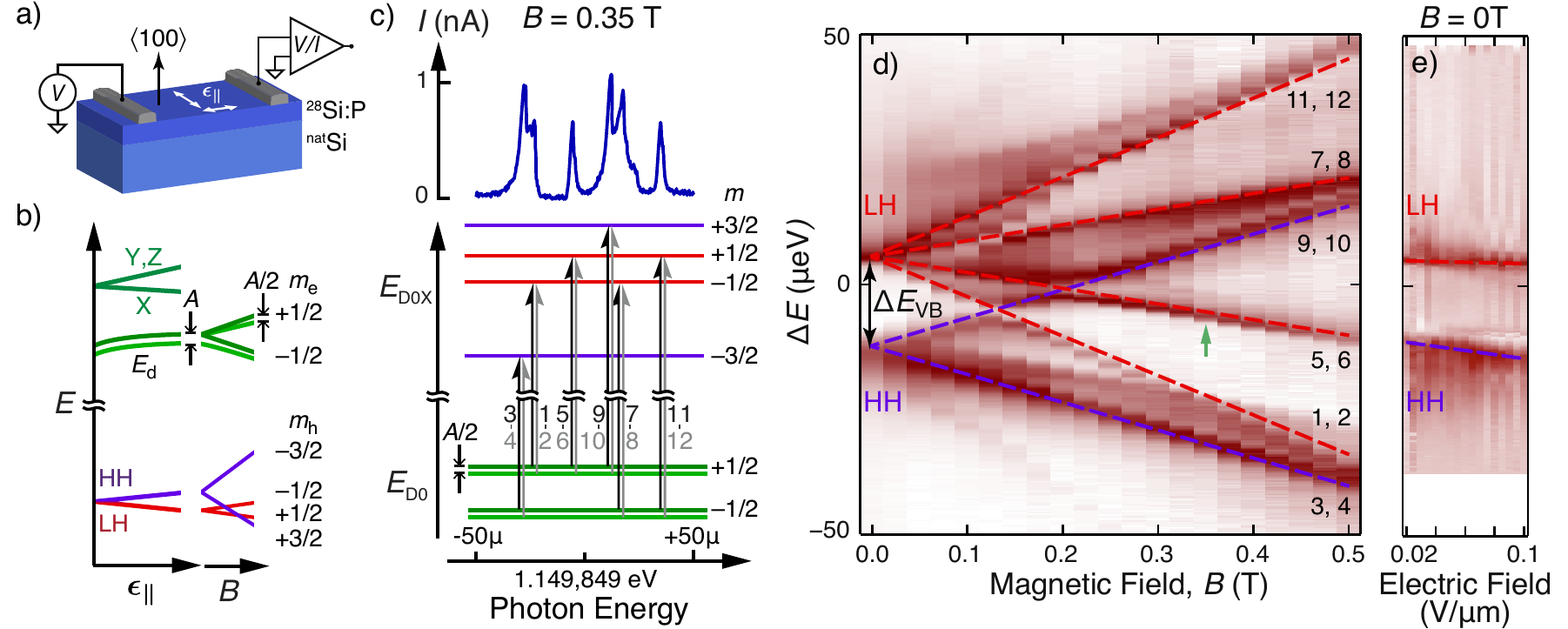}
	\caption{\label{f:1}{
Electrically detected D$\rm ^0$X spectroscopy in a silicon device. (a) Schematic of the silicon device built on a $^{31}$P-doped $^{28}$Si epitaxial layer on an undoped natural Si substrate, with crystallographic and in-plane strain ($\epsilon_{||}$) orientations indicated. (b) Energy shifts of the silicon conduction band vallies (X,Y and Z), donor ground state ($E_d$), and light hole (LH) and heavy hole (HH) valence bands due to $\epsilon_{||}$ and the applied magnetic field ($B$). (c) Corresponding allowed optical transitions for D$\rm ^0$X formation with $\Delta m\:=\:0, \pm1$ at $B\:=\:$0.35$\:$T (lower panel), and the measured spectrum from the silicon device (upper panel). The labels for the optical transitions follow the conventional numbering, see Ref~[\citen{steger12}]. (d) Magnetic field dependence of the D$^0$X spectrum with dashed lines showing theoretical fits based on the extracted $g-$factors and $\epsilon_{||}$. The green arrow indicates the optical transition and magnetic field used in subsequent measurements. (e) The electric field dependence of D$^0$X spectrum at $B\:=\:$0$\:$T shows the Stark shifts of the LH and HH bands. Dashed lines are linear fits to the LH and HH peak positions.}}
\end{figure}

The D$^0$X Auger recombination process has recently been used in conjunction with contact-less capacitive 
schemes in bulk silicon for the detection of nuclear spin coherence times in highly enriched \sitweight ~[\citen{steger12, saeedi13}]. 
We apply this technique for electrical detection of D$\rm ^0$X spectroscopy via direct transport measurements through devices built on epitaxially grown \sitweight\ doped with \ptone\ at $10^{15}\:$cm$^{-3}$ (see \Fig{f:1}). The \sitweight\ epitaxial layer has a built-in biaxial strain due to its lattice mismatch with the undoped substrate of natural isotope abundance\cite{yang08}. The presence of strain modifies the local bandgap surrounding the donors by lifting the degeneracy of the valence and conduction band edges, and consequently shifts the donor binding energies through valley repopulation\cite{wilson61}.
The states are further Zeeman split under an applied magnetic field, leading to six pairs of dipole-allowed transitions ($\Delta m\:=\:0,\pm1$), which we observe by monitoring the current through the device as the laser wavelength is swept. Due to the strain distribution in the epitaxial material and local strains induced by the electrical contacts, we do not resolve the hyperfine splitting of the \ptone\ donors in this silicon device. 
By mapping out the magnetic field dependence of the \dox\ spectrum we obtain a complete picture of the valence band light hole (LH), heavy hole (HH) and Zeeman splittings, where the measured zero field splitting $\Delta E_{\rm VB}\:=\:$19$\:\mu$eV is due to the $\epsilon_{||}\:=\:$+2.4$\times$10$^{-6}$ (tensile) biaxial strain-induced LH-HH splitting. 
Assuming an isotropic $g$-factor of $g_d\:=\:1.9985$ for the \ptone\ donor electrons\cite{feher59}, we find the D$\rm ^0$X hole-state  $g$-factors to be $g_{LH}\:=\:$0.86 and $g_{HH}\:=\:$1.33 in this magnetic field orientation ($B \parallel\:\langle100\rangle$), in good agreement with earlier measurements~[\citen{kaminskii80}] (see Supplementary Information).
Our device geometry allows us to study the effect of the LH-HH splitting as a function of electric field (\Fig{f:1}(e)), yielding a linear Stark shift parameter of 2$p_8\:=\:$33$\pm\:$7$\:\mu$eV/(V/$\:\mu$m), or $p_8\:=\:$0.8~Debye, similar to acceptor states in silicon\cite{kopf92}. Carrier injection from the electrical contacts prohibits measurements at larger electric fields.

\begin{figure}
	\centering
	\includegraphics[width=16.0cm]{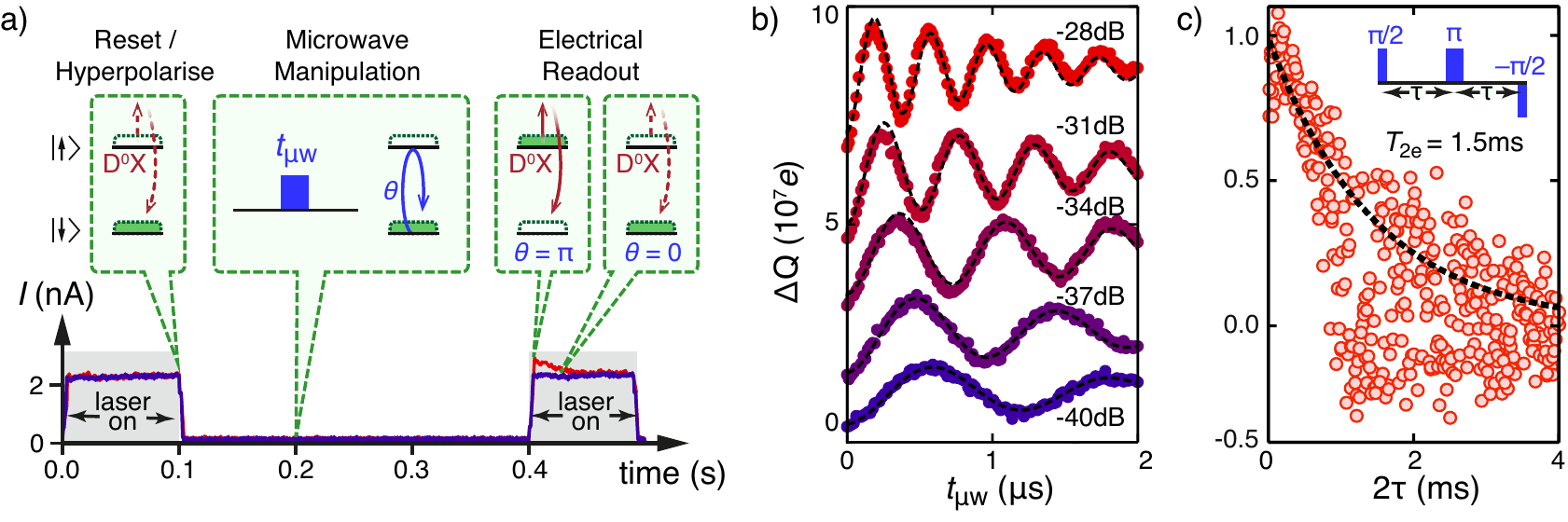}
	\caption{\label{f:2}{Electrically detected pulsed electron spin resonance utilising D$^0$X. (a) Laser pulses (grey) consist of the D$^0$X laser tuned to transitions 5 and 6 (unresolved in this sample), and a weak above-bandgap laser at 1047$\:$nm which expedites the donor re-neutralisation process. An initial laser pulse hyperpolarises the donor electron spins into the $|\downarrow\rangle$ state, and is followed by a microwave pulse of duration $t_{\rm \mu w}$ corresponding to some rotation $\theta$ ($\theta=0$, purple trace, or $\theta=\pi$, red trace). The current transient during a second `readout' laser pulse is used to measure the electron spin population in the $|\uparrow\rangle$ state. Coherent control and electrical detection of donor states demonstrated by (b) Rabi oscillations (with the microwave power attenuation levels as indicated, offset for clarity), and (c) a $T_2$ measurement where the microwave pulse is replaced with a Hahn echo sequence. Dashed lines are fits to the experimental data and all experiments are performed on resonance with the $m_I\:=\:-1/2$ ESR transition.}}
\end{figure}

To demonstrate electrical detection of electron spin resonance using \dox\, we set the magnetic field to $B\:\approx\:$0.35$\:$T and tuned our laser on resonance with transitions 5-6, as shown in \Fig{f:1}(d). 
This optical excitation drives the spin-selective ionisation of spin-up electrons, after some time ($\sim100$~ms) leaving the donor electrons hyperpolarised into the spin-down state\cite{yang09}.
The laser excitation is turned off to allow coherent control of the donor spins via applied microwave pulses, and then turned on again for readout. \Fig{f:2}(a) shows the measured current transients with, and without, a microwave $\pi$ pulse applied, illustrating how the difference in the integrated signals between the two is a measure of the donor spin $z$-projection. Using this method we measure electron spin Rabi oscillations (\Fig{f:2}(b)), demonstrating coherent manipulation and electrical detection of the donor spin states and yielding an ensemble dephasing time of $T_2^*\:=\:$2$\:\mu$s caused by the strain distribution in the sample. 

We measure the electron spin coherence time by implementing a two-pulse Hahn echo sequence with an additional pulse used to project the coherence into the population of the spin eigenstate (\Fig{f:2}(c)).
The fluctuations in signal amplitude after 1$\:$ms are due to the presence of instrumental magnetic field noise so that only the period $2\tau\leq1$~ms is fit to extract the spin coherence time. 
The measured value of $T_2\:=\:1.5\:$ms is in good agreement with bulk spin resonance measurements for samples under similar dopant concentrations and isotopic purity\cite{tyryshkin12}, and reflects the advantage of utilising spin-selective optical transitions for electrical readout, as the donor spin coherence is no longer inherently limited by nearby sources of decoherence (e.g. $T_2\:\approx\:1\:\mu$s for spin-dependent recombination with interface paramagnetic defects\cite{paik10}).

\begin{figure}
	\centering
	\includegraphics[width=8cm]{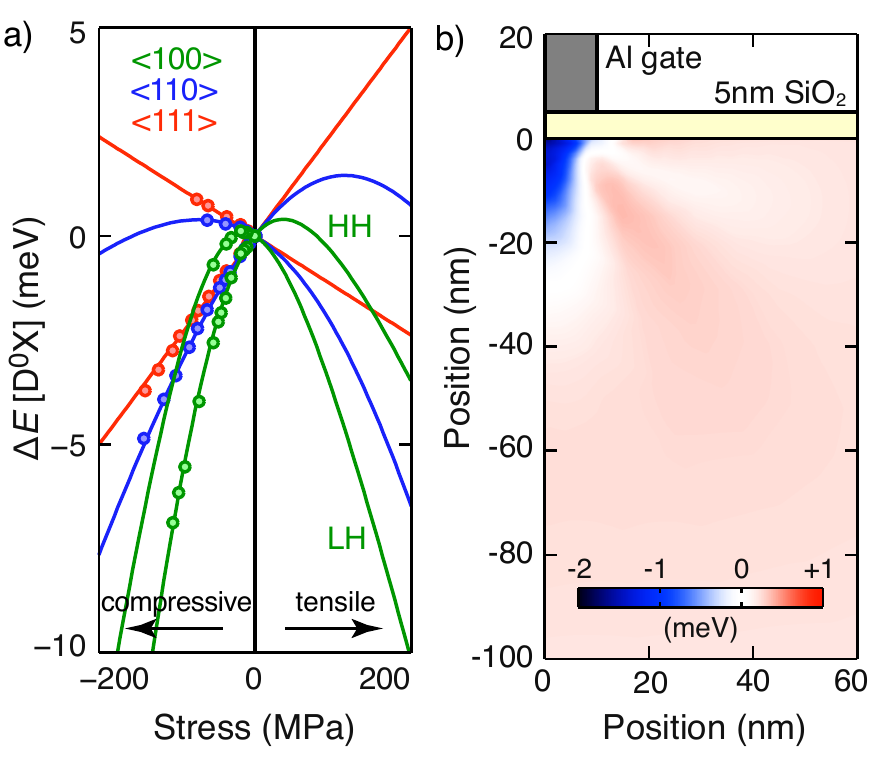}
	\caption{\label{f:4}{Strain-induced shifts in \dox\ transition energies, $\Delta E[\mathrm{D^0X}]$. (a) Uniaxial stress applied along the major crystallographic directions: $\langle100\rangle$ (green), $\langle110\rangle$ (blue) and $\langle111\rangle$ (red). The lines are results from our calculation, and circles are data taken from Ref~[\citen{thewalt78}]. (b) Calculated $\Delta E[\mathrm{D^0X}]$ due to thermal expansion coefficient mismatch for \ptone\ donors in close proximity to Aluminium electrodes and under a 5~nm gate oxide at 4.2~K. Only the higher energy of the two strain-induced valence band branches for the transitions are shown. See Supplementary Information for details.}}
\end{figure}

Below we address additional considerations for extending these results to realise single donor spin detection utilising neutral donor bound excitons, in particular the effects of i) strain and electric fields in nanodevices and ii) off-resonance optical excitation on readout fidelity. 
As observed in our silicon device, the \dox\ transition energies are extremely sensitive to local strains in the silicon crystal. This is further exemplified by the fact that the \dox\ inhomogeneous linewidth differs widely in bulk-doped electronic grade silicon crystals: $\approx\:$100$\:\mu$eV for Czochralski silicon
, $\approx\:$5$\:\mu$eV for float-zone silicon and $\leq\:$0.2$\:\mu$eV for high-quality isotopically enriched $^{28}$Si~[\citen{yang09}]. 
Figure$\:$\ref{f:4}(a) shows the change in \dox\ transition energies, $\Delta{\rm E}$[\dox], for \ptone\ donors at zero magnetic field calculated from a single-particle perturbative model (see Supplementary Information for details) for uniaxial stresses up to 200$\:$MPa ($\approx$ 10$^{-3}$ strain), which are not uncommon in silicon nanodevices\cite{thorbeck14}. The two branches for each stress direction are due to the valence band splitting as we have observed in our silicon device.
Strain is inevitably present in qubit devices close to the silicon surface, where device-dependent strains can be introduced during the fabrication process, or due to the thermal expansion coefficient mismatch between gate or dielectric materials\cite{thorbeck14}.
This is illustrated in Figure$\:$\ref{f:4}(b), showing calculated $\Delta{\rm E}$[\dox] due to the thermal expansion coefficient mismatch problem for \ptone\ donors in close proximity to aluminium gate electrodes. 
The presence of metallic gates and the insulating oxide strains the silicon substrate considerably, and the uncertainty in dopant positioning can cause a large detuning of the transition energies that are orders of magnitude greater than the spin splitting, limiting the ability to address arrays of single dopant devices in a scaled qubit architecture.
This sensitivity to strain is more severe for D$\rm ^0$X than for optical transitions of erbium ions in silicon\cite{yin13}, where only core shell electronic levels are involved.  

The electric fields present in silicon nanodevices will also shift the \dox\ energy, as we have already shown above through the Stark splitting of the hole states. However, if the field is particularly strong, it is expected to additionally reduce the bound exciton lifetime in an analogous process to field ionisation of donors, but taking into account the much lower binding \dox\ energy of 5~meV.
Therefore, both strain and electric fields in realistic single dopant devices must be carefully considered and controlled for the successful implementation of D$^0$X-based spin readout. 

\begin{figure}
	\centering
	\includegraphics[width=8.0cm]{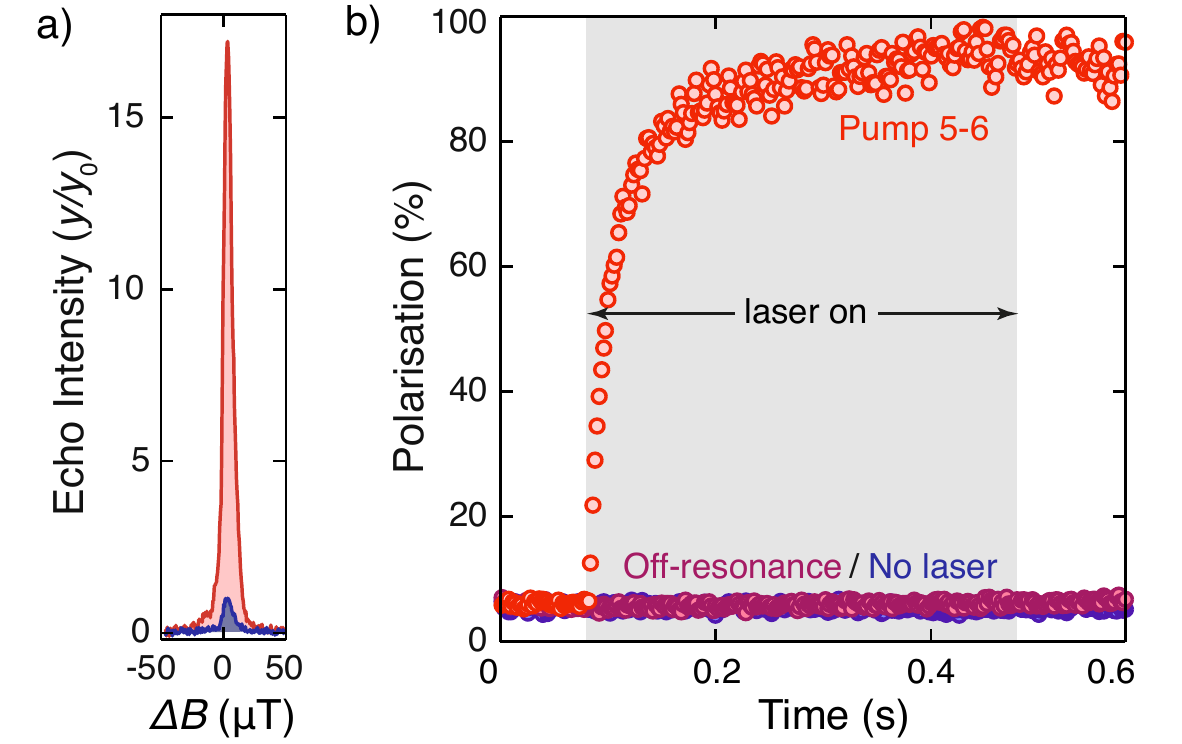}
	\caption{\label{f:3}{ESR-detection hyper polarisation under D$\rm ^0$X laser excitation. (a) Electron spin echo-detected field sweep of the $m_I\:$=$\:$-1/2 hyperfine line of $^{31}$P donors in a \sitweight\ crystal, measured at 4.2$\:$K in thermal equilibrium (blue), and with the laser tuned to transitions 5-6 (red), showing a signal enhancement by a factor of $\sim18$. (b) The dynamics of the hyperpolarisation process is studied with the D$^0$X laser tuned to transitions 5-6 (red), tuned off resonance by 3$\mu$eV (purple), and measured with the laser turned off completely (blue). The lack of response when the laser is off resonance indicates non-resonant ionisation processes are negligible here.}}
\end{figure}

Two important factors impacting electron spin readout fidelity using this approach are the ability to detect the resonantly-formed ionised donor state (e.g.\ using a charge sensor) and false readings arising from non-resonant ionisation events. 
The former is unlikely to be limiting as bound excitons can be resonantly generated at a much faster rate than typical spin relaxation times ($\approx\:$85$\:$MHz  in our experimental set up, see Supplementary Information), and once ionised, the dopant neutralisation time is in the order of 10$\:$ms (depending strongly on the concentration of free carriers) which is ample time for charge state detection. 
On the other hand, non-resonant ionisation of the donor state can be caused by off-resonance photons directly ionising neutral donors, or creating free excitons which subsequently recombine at neutral donor sites (we note, however, the energy required for free exciton formation is approximately 5$\:$meV higher than resonant \dox\ transitions in bulk silicon). We examine the role of non-resonant irradiation in more detail below by observing its effect on bulk electron spin resonance (ESR) measurements. 
We first show the electron spin hyperpolarisation through the ESR signal enhancement seen for an echo-detected field sweep, with and without a laser pulse tuned to the 5-6 \dox\ transition (\Fig{f:3}(a)).
The dynamics of this process can be seen by mapping out the echo intensity at different times during the laser pulsing sequence, as shown in \Fig{f:3}(b). 
The echo intensities are normalised with respect to the thermal equilibrium signal, and when the laser pulse is on resonance, the maximum electron spin polarisation achieved is close to 100$\%$ (see Supplementary Information). 
Conversely, when the laser is detuned away from resonance by 3$\:\mu$eV, the echo intensity remains constant and identical to the thermal equilibrium measurement. This demonstrates that off-resonant ionisation is negligible in dilutely doped substrates (as both direct ionisation and free exciton-donor recombination would diminish the echo intensity) and hence is unlikely to pose a major limit on electron spin readout fidelity. Nevertheless, if non-resonant ionisation is found to affect readout fidelity to some degree, the donor nuclear spin could be used as an ancilla for performing repetitive measurements\cite{jiang09}. 

A hybrid optical-electrical single donor detection scheme utilising donor bound excitons coupled to a quantum point contact has been previously proposed\cite{sleiter10}, however, the uncertainty in local strain and relatively large electric fields present in metal-oxide-semiconductor based architectures will make measuring bound excitons difficult. 
While D$\rm ^0$X detection can conversely be used as an extremely sensitive probe to quantify strain and electric fields of silicon nanostructures and devices in the atomistic scale, reducing these perturbative field distributions will be crucial for implementing hybrid optical-electrical detection for large arrays of qubit devices on a single chip.
Therefore, an optimally designed readout device should have both the strain and electric fields carefully controlled (and minimised) in the vicinity of the dopant. For instance, epitaxially-grown\cite{fuhrer09} or nanowire single electron transistors operating at liquid helium temperatures can be used as the charge detector, and in this hybrid spin detection scheme the need for dilution refrigerators can be completely alleviated, opening the door for exploiting ultra-long coherent donor spins for practical quantum technological applications.

We thank A.M. Tyryshkin for useful discussions. This research is supported by the EPSRC through the Materials World Network (EP/I035536/1) and UNDEDD project (EP/K025945/1) as well as by the European Research Council under the European Community's Seventh Framework Programme (FP7/2007-2013)/ERC grant agreement No.\ 279781. Work at Princeton was supported by NSF through Materials World Network (DMR-1107606) and through the Princeton MRSEC (DMR-0819860).  C.C.L. is supported by the Royal Commission for the Exhibition of 1851. J.J.L.M. is supported by the Royal Society.

\end{document}